\documentclass[showpacs,prl,twocolumn,groupedaddress,aps]{revtex4}
\usepackage{graphicx, epsfig}

\newcommand{\be}{\begin{equation}}
\newcommand{\ee}{\end{equation}}
\newcommand{\bea}{\begin{eqnarray}}
\newcommand{\eea}{\end{eqnarray}}

\begin{document}

\title{The Arrow of Time Forbids a Positive Cosmological Constant $\Lambda$}

\author{ L.~Mersini-Houghton}
\email[]{mersini@physics.unc.edu}
\affiliation{Department of Physics and Astronomy, UNC-Chapel Hill, NC, 27599-3255, USA}

\date{\today}
 
\begin{abstract} 
Motivated by the mounting evidence for dark energy, here we explore the consequences of a fundamental cosmological constant $\Lambda$ for our universe. We show that when the gravitational entropy of a pure DeSitter state ultimately wins over matter, then the thermodynamic arrow of time in our universe must reverse in scales of order a Hubble time. We find that due to the dynamics of gravity and nonlocal entanglement, a finite size system such as a DeSitter patch with horizon size $H_0^{-1}$ has a finite lifetime $\Delta t$. This phenomenon arises from the dynamic gravitational instabilities that develop during a DeSitter epoch and turn catastrophic. A reversed arrow of time is clearly in disagreement with observations. Thus we are led to conclude: Nature forbids a fundamental $\Lambda$. Or else general relativity must be modified in the IR regime when $\Lambda$ dominates the expansion of the Universe.
\end{abstract}


\maketitle
We live in a world that has an arrow of time. The second law of
thermodynamics states that the Universe must have started in an extremely
low entropy state. Low entropy states are but a small subset of the general
phase space for the initial conditions. Penrose argued that this seems to make the choice of
our Initial Conditions very special indeed. The loophole in the above argument lies on eliminating the possibility of
some dynamic selection being at work. Statistical
mechanics estimates, often used in literature, can quantify the Penrose
statement: the probability to have an initial patch inflate at some high
energy scale $\Lambda_i$ goes as $P \simeq e^{S_i}$ where the entropy
$S_i =3/\Lambda_{i}$. This expression indicates that a GUT scale inflating patch like ours, is the most special and unlikely event to have
started the universe, as likely as 1 part in $10^{10^{123}}$. Yet without
this low entropy we can't explain the observed arrow of time.

In \cite{richlaura1} we proposed a new approach, based on the quantum dynamics of the gravitational
degrees of freedom (D.o.F), for the investigation of the Initial Conditions. We showed how the dynamics of the combined,
matter +gravity system, is out-of equilibrium and how this breaks ergodicity.
The compression of
the phase space into only a small region of initial conditions
corresponding to low entropy and high energy inflation resulted in the emergence of a new superselection rule for Initial Conditions. The superselection of high energy patches from the non-ergodic phase space
helped us understand why we have to start from extremely low entropies, thereby with an
arrow of time.

Let us briefly highlight the dynamics of the combined system below as it becomes relevant again for the late time accelerating Universe.
The basic mechanism that picks out the high energy 'corner' in the phase space
is the following: gravitational D.o.F, here corresponding to the degrees of freedom on the DeSitter horizon $ H_0 \simeq \sqrt{\Lambda}$, comprise a 'negative heat capacity'
system, therefore gravity tries to reach equlibrium by tending the
spacetime to infinity; matter is a 'positive heat capacity' system and thus
goes towards equilibrium by collapsing; the dynamics of the combined
system, matter +gravitational D.o.F, is an out-of equlibrium system with
two opposing tendencies 'fighting' to expand or collapse the patch, which, depending upon 
which one wins, can result in a collapsing initial patch, thus creating irrelevant
Initial Conditions for starting a universe or, if gravity 'wins' in an
expanding initial patch which result in a physically relevant Initial
Condition that gives birth to a universe. We belong in the
latter, the 'survivor universes' case. The entropy of the 'survivor
universes' given by the Euclidean action of the system, is modified by gravity and in fact even
less than its equilibrium expression, due to the out-of equilibrium
dynamics. Entropy is given roughly by log of the volume in phase space. The modified entropy is then a direct consequence of the 
reduction of the relevant phase space to a small 'corner' of its initial volume that contains the 'survivor
universes' only.

It is possible that our Universe may be going through inflation again at
late times. If true, then the dynamics of the gravitational degrees of freedom is expected to play a major role again at late times. Within the framework of our proposal that becomes relevant in this case as here, we would like to ask: What is
the ultimate state of such $\Lambda$-universe and can it achieve equilibrium?

Current astrophysical observations provide convincing evidence that our
universe is indeed accelerating again at recent times, this time at an energy scale
$\Lambda \simeq (10^{-3} eV)^4$. Determining whether a pure vacuum energy
$\Lambda$ or some dynamic mechanism is at work is of paramount importance
and intrinsically related to understanding and predicting the future
evolution of our universe. 
But, it is well known that cosmological observables are observer dependent. Thus
we need to define our observer. A key point to keep in mind is that in
order to {\it meaningfully} ask questions about the selection of our
Initial and Boundary Conditions, quantum numbers or constants of nature, we need to rely
on a super-observer, defined as the one that lives in the multiverse of the
landscape or equivalently in the allowed relevant part of the phase space.
In short we need an out-of-our box observer in order to compare posssible
choices\cite{richlaura2,land1,laura}.
A local internal observer, defined as the one bound to our casual patch ('box'), is
always surrounded by the horizon thus (s)he can
never observe or be aware of the existence of other patches outside the
horizon. It is impossible for the internal observer to
ask questions about the selection of our initial and final conditions and
our constants of nature because (s)he is forbidden to ever know that
possibilities other the universe (s)he observes, may exist.
Using the higher multipoles as the super-observer, in this letter we explore the
implications of the proposal in \cite{richlaura1} for the late time acceleration of the
universe.

We show below how the final destiny of a DeSitter universe is to become quantum again, roughly in times of order Hubble time, thereby reversing its thermodynamic arrow of time at the turning point of its classical trajectory. We will refer to the transition to a quantum state as recoherence of the wavepacket. The details of the quantum state near the turning point depend on the boundary conditions chosen for the wavefunction which are important for the puzzle of the agreement between a thermodynamic and a cosmological arrow of time.We do not embark upon this puzzle here.The key point in this work is the fact that the Universe becomes quantum while it is trying to reach the thermal equilibrium of a pure classical DeSitter state and thus it never reaches this equilibrium. We investigate how the universe transits into the quantum state when it approaches a pure DeSitter state and how this transition affects the thermodynamic arrow of time.  
We argue on these basis that within the framework of general relativity as the
theory of gravity, nature forbids the existence of Desitter spaces, i.e of
a pure fundamental constant $\Lambda$. We then speculate in the discussion
section that that leaves us either with the possibility of a temporary
bleep of acceleration due to some dynamic \cite{quintessence} or
transplanckian \cite{lauratp} mechanism or, to the conclusion: {\it if we must
have $\Lambda$ then we must have a new theory of gravity in the IR regime
which replaces general relativity} \cite{mond}.

\section{Thermodynamic Arrow of Time Does Not Allow DeSitter Space}

Suppose we do have a new fundamental scale at low energies in our Universe,
the DeSitter horizon $H_0 \simeq \sqrt{3/\Lambda}$, given by a pure vacuum
energy $\Lambda \simeq (10^{-3} eV)^4$. Observations will soon pin down the equation of state $w$ for dark energy, thus distinguishing
between a pure $\Lambda$ and a dynamical mechanism. Observationally we also
know and agree that we have an arrow of time, pointing from past to future,
determined by the direction of the entropy growth. What does the constant horizon entropy
of the final state of the $\Lambda$CDM universe imply for
the arrow of time and ultimately for the DeSitter universe?

Based on the assumption that dark energy is a fundamental $\Lambda$, it seems that we
ultimately must end in a DeSitter space. Gravity ultimately wins
\cite{gibbonshawking} for the following reason:in contrast to the entropy of DeSitter space which remains eternally a finite constant, not only does matter dilute with time, \footnote{(e.g. in Black Holes by radiating away)} but,  as we accelerate, parts of
our spacetime lose causal contact with each other, swallowing matter with
them. Ad infinitum we lose all the matter from our universe in this manner
and the entropy contribution associated with it. Ultimately the only
entropy component left is the gravitational one associated with the
DeSitter horizon. Unlike Black Holes, a DeSitter horizon does not evaporate away. Matter provides the environment that ensured decoherence, thus the Gaussian suppression of width ${\Omega_R}^{-1}$ in the off-diagonal elements of the density matrix $\rho$, (\cite{starobinski,kiefer,kiefer2}, see also Eq.\ref{eq:density1}), which describe how fast our (patch) vacua $\phi$ decoheres and becomes classical,
$\rho \simeq e^{-\Omega_R a^4 (\phi - \phi')^2}$. Decoherence and the appearance of a classical world comes about from the backreaction of the environment on our branch of the wavefunction $\Psi(a,\phi)$. Losing causal contact with matter
seems to imply that we reach thermal equilibrium by evolving in a pure DeSitter state since we lose the environment and the coarse-grained entropy growth provided by it. This reasoning implies that the energy level in phase space for our DeSitter universe should broaden. Since the DeSitter
horizon $H_{0}$ classically is a constant then at the classical level so is its
entropy, constant and finite. But the thermodynamic arrow of time is determined by the entropy growth. At the classical level, it is not clear at all what happens to the arrow of time during an eternal DeSitter epoch when its classical entropy remains a constant. These paradoxes indicate that we must search for an answer not at the classical level but by investigating the quantum dynamics of the gravitational degrees of freedom of the system.

We showed in \cite{richlaura1} that during an inflationary phase higher multipoles develop gravitational Jeans
-like instabilities which illustrates the importance of the dynamics of gravitational degrees of freedom. The entropy of the universe is lowered in the combined
matter+gravity inflating system. At early time, only patches inflating at high energies $\Lambda_i$, survived this instability.
We discussed in \cite{richlaura1, richlaura2} how the mixing in the initial
state must survive at late times,due to unitarity, (see also \cite{shenker}). The reduced density matrix
obtained by integrating out the environment and superHubble matter modes,
gave a scale of mixing and decoherence for our patch at initial times. It also provided us with
the evolution equation of this mixing/coherence length at late times namely, $[\hat H, \rho] = i\frac{d\rho}{dt}$ .


The initial decoherence width described how fast the quantum to
classical transition for our universe occurred at early times. The fact that the
density matrix evolves with time $[H, \rho_r] \ne 0$,
\cite{richlaura1,richlaura2} illustrates why, with gravity switched on, we should not be tempted to use
equlibrium expression like $P\simeq e^S_E$ again, and why we can not use
causal patch observers for investigating this problem. 
The universe is in a mixed state and out-of equilibrium even at late times. How can it evolve to a pure state to reach
equilibrium?

Classically, the ultimate destiny of the $\Lambda$ universe
is a cold DeSitter space in thermal equilibrium, with finite final entropy $S \simeq
3/ \Lambda$. Can an ultimate pure state be allowed for the DeSitter space?  The environment
in our investigation of the dynamics of matter
+gravity D.o.F., is made of the matter modes within our casual patch as
well as the backreaction of higher multipoles with superHubble wavelengths.
By losing matter inside our casual patch means we are losing part of the
environment, which is very important in providing the Gaussian suppression
for the off-diagonal elements of $\rho_r$ that ensure decoherence and a
classical world. Gravity winning over matter seems to imply that we are
evolving into a pure state, rather than a mixed state on the
landscape phase space. {\it But, if we assume a unitary evolution, we can not
evolve in a pure state}. Perhaps, even with matter gone,some degree of mixing must remain. This entanglement can come only from interaction with superhorizon modes, i.e. from 'out-of-the box'. In fact it does. In the DeSitter stage, the environment comes from the entanglement and interaction of our DeSitter patch with the higher multipoles\cite{richlaura1}. The dynamics of this mixed state is highly nontrivial and we set to explore it below.

Many authors have discussed the discrepancies and apparent paradoxes that
arise from a finite DeSitter entropy in the final state of the universe. In
\cite{witten,susskind,causalpatch,eqdesitter} authors point out that a finite entropy implies a
finite size Hilbert space which is incompatible with DeSitter symmetries
$SO(d,1)$. Meanwhile in \cite{bousso} by geometric arguments, authors
realized that DeSitter space may be unstable, in accord with our results in
\cite{richlaura1,richlaura2}.  Authors of \cite{susskind} then argued that
perhaps DeSitter state is metastable on the landscape and it either goes
through a Poincarre cycle or tunnels to lower energy vacua. The decay time
for tunneling to lower states $T_f$ seems shorter than the recurrence time
$T_r$, $T_f \simeq e^{S_o - S_f} << T_r \simeq e^S$, where $S_o$ is the vacua entropy and $S_f$ the fluctuation entropy that
would take the state to the top of the potential and allow it to roll the
other way into the lower energy vacua. This picture fits well with eternal
inflation where many pocket universes at different energies are created and
transition among them possible. The possiblity that a DeSitter state is not
eternal but rather metastable due to tunneling to lower energy states, is
certainly plausible.

But, even if it sounds unlikely, what if $\Lambda$ is a globally
fundamental scale, i.e suppose that this DeSitter state has no other
lower state to tunnel to? What then?

Let us show that even without tunneling, a DeSitter state is not
just metastable, but gravitationally it is a catastrophically unstable state. We show here that even before it becomes metastable from tunneling,
the {\it growth of gravitational instabilities} force the DeSitter patch to reverse its arrow of time by hitting a turning point in its trajectory in phase space. We show this phenomenon occurs in time scales
 of order roughly horizon scale, $T_{crunch}\simeq \sqrt{\Lambda}$ which is exponentially shorter that tunneling time $T_f$.This result would be in huge disagreement with observations. We do not observe a reversal of the arrow of time in our DeSitter universe, in time scales of the order of the age of our universe.
\subsection{Big Crunch Through Recoherence or a New Exclusion Principle?}

 Suppose we are in the far future where we have lost almost all
matter contact from our casual patch. As we discussed above, the only environment we are left with now comes from
the backreaction of the superHubble matter modes, i.e matter fluctuations of order
DeSitter horizon $H_0$ or larger. We would like to know the evolution of the wavefunction for our universe, and the evolution of the density matrix. The latter will describe the effect of the interaction of superhorizon modes on our DeSitter geometry.Proceeding as in \cite{richlaura1}, let us assume that within the WKB approximation, the wavefunction is given by, $\Psi[a,f_n,\phi]\simeq e^S_E
$, $\Psi=\Psi_o \Pi_n \psi_n$, $S \simeq S_0 + \Sigma_n S_n \frac{1}{2}f_n ^{2}$ where fluctuations $f_n$ correspond to an expansion around zero of the matter field $\phi =\phi(0) + \Sigma_n f_n(a) Q_n(x)$, which can be written as

\be
\label{eq:wavepert}
\Psi \sim \Psi_0 (a, \phi_0) \prod_n \psi_n (a, \phi_0, f_n).
\ee

The Wheeler-DeWitt Master Equation, that includes the backreaction $\hat{H_n}$ of matter higher multipoles $f_n$ \footnote{(obtained by the expansion around zero of the landscape minisuperspace field $\phi$, see \cite{richlaura1,richlaura2})} is obtained from the quantized Hamiltonian 
\be
\label{eq:quantumH}
\hat{H} = \hat{H}_0 + \sum_n {\hat{H}}_n
\ee
 acting on the wavefunction, \cite{hh,kiefer2}

\be
\label{eq:master}
\hat{H}_0 \Psi (a, \phi_0) = \left(-\sum_n \langle  \hat{H}_n\rangle\right) \Psi (a, \phi_0),
\ee
where the angular brackets denote expectation values in the wavefunction $\psi_n$ and $\hat{H_n}$ denotes the backreaction correction from interaction with the superhorizon fluctuations $f_n$. The potential term of the backreaction corrections during a nearly DeSitter phase \cite{richlaura1, kiefer2},calculated from the inhomogenous matter perturbations in \cite {hh},to second order, is
$$
U_{\pm} (\alpha, \phi) \sim e^{6\alpha} \left[\frac{n^2-1}{2}e^{-2\alpha}\pm \frac{m^2}{2}\right].
$$, with $m^2 \simeq \sqrt \Lambda$ at present and $U_{\pm}$ sign corresponding to the interaction coming from subHubble(+) or superhorizon(-) multipoles respectively, as summarized in the Appendix. During the DeSitter stage the backreaction energy density $<U f_n^{2}>$ becomes negative, $U=U_{-}$, since our environment contains only the modes with wavenumber $n \le aH_0$ i.e with superhorizon wavelengths\cite{kiefer,richlaura1}. The index $0$ refers to present day values and $H_0 \simeq \sqrt{\Lambda}$ is the present DeSitter Hubble constant, with $a$ the scale factor. This negative contribution yields an oscillatory correction to the action and it gives rise to tachyonic fluctuations in Eq.\ref{eq:master}. The backreaction hamiltonian, during the DeSitter stage with curvature $\Lambda$, is roughly

\be
\label{eq:nham}
 2 e^{3\alpha} \hat{H_n} = -\frac{\partial^2}{\partial f_n^2} + e^{6 \alpha} \left( - \sqrt{\Lambda} + e^{-2 \alpha} \left(n^2-1\right)\right) f_n^2,
\ee

with $\alpha = Log[a]$, $\hat{H_n}\simeq \Sigma_{n} U_n f_n^{2}$ where matter fluctuations $<f_n^{2} >=
O (H_0)$. During the matter dominated phase the superhorizon modes are nearly frozen thus their kinetic term contribution can be ignored. During the DeSitter phase with vacuum energy $\Lambda$, our environment consisting of superhorizon matter fluctuations, contributes a negative backreaction, $f_n^{2} U_n = - (\sqrt{\Lambda} - n^2/a^2) f_n^{2}$, as can be seen from the expression that $U_n <0$ is negative for the superhorizon higher multipoles $n \le aH_0$, \cite{kiefer,richlaura1}. This negative contribution drives the the gravitational instability of the massive superhorizon size modes $f_n$, that is, during the DeSitter stage $f_n$ modes start growing exponentially with time, (see the Appendix and \cite{richlaura1} for the $f_n$ equation of motion).
 
The correction term $\Psi_n$ to the wavefunction satisfies the following equation
\be
i \frac{\partial{\Psi_n}}{\partial{t}} = \hat{H_n}\Psi_n
\ee
which for the modes $n \le aH_0$ with $H_0$ our present DeSitter horizon, and $U_n <0$ gives a solution of the form, $\Psi_n \simeq e^{-\frac{1}{2}f_n^{2} S_n(a)}$ where, the WKB 2nd order correction term to the action is $S_n \simeq i\int a^3 U_n \frac{dln(a)}{H}$. Since $U_n <0$ and $S_n$ is not real in this regime, then this correction breaks the finitiness condition \footnote{We can treat the contribution of each mode $f_n$ separately since they are decoupled from each other.}.

Perhaps a better way to see the role of the superhorizon size nonlocal entanglement of our DeSitter patch with $f_n$, or equivalently in the phase space description, the effect of the backreaction term on the classical path of the wavefunction is by
integrating out the Friedman-like equation which includes the entanglement term. This equation is obtained from the generalized
hamiltonian in the Master Equation Eq.\ref{eq:quantumH},  which contains the backreaction energy, $\Sigma_n \hat{H_n}$

\be
\label{eq:turningpt}
3M_{p}^2 \dot a^2 - \Lambda a^2 + a H_0^3 =0
\ee

where $\Lambda = 3M_{p}^2 H_0^{2}$. The last term comes from the backreaction of the higher multipoles $k=n/a \le H_0$ thus it is negative during the DeSitter stage as explained above. A rough estimate is obtained by summing over all the superhorizon modes, $\Sigma_{aH_0}^{0} <\hat{H_n}>$, where $\Sigma_{k=n/a=H_o}^{0} (n/a)^2 dn f_n^2 \simeq a H_0^{3}$.(Below we take $8\pi G_N =1$ unless explicitly stated). From Eq.\ref{eq:turningpt} we can see that our classical trajectory reaches a turning point in phase space, $\dot a_* =0$,
during a finite time $a_*$ given by

\be
a_{*}^2 \Lambda = a_* H_{0}^3, \\  a_* =\sqrt{\Lambda}
\ee 

This means that due to the gravitational instability developed by the
tachyonic superHubble modes, $U_n <0$, our DeSitter space goes through a
crunch in real spacetime or a turning point in phase space in a time given
by the condition $a_* = O(H_o)$. Solutions to Eqn.\ref{eq:turningpt} allow us to estimate the time, $t$, that the transition takes to go, from the regime of a classical large $\Lambda$ dominated DeSitter universe with scale factor $a$ and Hubble expansion, $H_0$, to the quantum regime of the Universe as it approaches the crunching turning point, where the scale factor,$a\rightarrow a_*$ becomes very small and the Hubble expansion rate, $\dot{a}\rightarrow \dot{a_{*}}=0$ approaches zero. This solution is

\be
a(t)=(\frac{H_0^{3}}{\Lambda}) Cosh^2[\frac{1}{2} H_{0} (t-t_{*}) ]
\label{eq:crunchtime}
\ee
Clearly when $t\rightarrow t_{*}$ we have $a\rightarrow a_*$, i.the universe's scale factor is reducing to a very small size and the turning point is approached at time $t_{*}$. Prior to the instabilities growth, in the regime where the universe is a classical $\Lambda$ dominated DeSitter space i.e. for $t \ll t_{*}$, we see that the scale factor $a$ is nearly exponentially large and, corresponding to that of a large DeSitter space. {\it We can readily estimate from our solution, Eq.\ref{eq:crunchtime} that the time $\Delta t = (t-t_{*})$ the Universe takes to transit from a large classical regime with size given by the Hubble radius,$a=r_{H_{0}}$, to a small quantum regime $a=a_{*}$ is $\Delta t \simeq \frac{122}{H_0}$, roughly about $122$ times the DeSitter Hubble time}.This is an amazing result, nothing short of an uncertainty principle! In this work, by taking into account the quantum dynamics of gravity, Eqn.(7), leads us to an uncertainty principle for our quantum cosmological 'intrinsic clocks' given by $a(t)$ for the DeSitter system with size $H_0^{-1}$ namely, $a(t) H_0^{-1}\simeq l_p^{2}$ where Planck length $l_p$ is unity in Planck dimensions. Using the solution for $a(t)$ in Eqn.(\ref{eq:crunchtime}) this result can be translated into an uncertainly principle for the physical time $t$ of the DeSitter patch, given by  $\Delta t H_0 = \Delta N \simeq Log[M_p^{4}/\Lambda]$ where $\Delta N$ is the number of efoldings that have crossed the horizon since the inflationary time, i.e. defined by $a=e^N$. It is reassuring that although DeSitter space is eternal at the classical level,our findings for a catastrophic collapse of DeSitter space at the quantum level during the time interval $\Delta t$ due to nonlocal entanglement and the dynamics of gravity, are consistent with the uncertainty principle. Furthermore, the interpretration we offer above for the nonlocal entanglement of the superhorizon modes $f_n$ performing a quantum measurement on the DeSitter patch is now justified by this uncertainty principle.
(Notice that in the above Eqn.\ref{eq:turningpt} we have not included the matter and radiation contribution to the energy density of the Universe, as we are interested here in investigating the transition regime when the Universe is becoming a DeSitter one,i.e from the time when $\Lambda$ dominates over the other components, and onwards. Subsequently the solution found for $a(t)$, Eq.\ref{eq:crunchtime}, reflect that fact through the nearly exponential time dependence, the $Cosh^2$ term.).

 Since our classical trajectory $\phi(a)$ goes through a
turning point induced by the gravitational instabilities of backreaction in the DeSitter epoch,$\dot{\Psi}=0$, it follows that after losing casual contact with the internal environment, the universe becomes quantum again as we approach the turning point in a time $a_*$. If observers bound to the causal patch survived while the universe is transiting to a quantum state, they would perceive the arrow of time as being reversed near the turning point. However notice that very close to the turning point, the WKB approximation breaks down, therefore the concept of time and any statement about what could be observed becomes very fuzzy. A better estimate as to when effects related to 'recoherence' display a significant effect in our universe, is obtained from the density matrix below, Eq.(11).


We would like to explore the evolution of the density matrix $\rho$ and the Gaussian suppression width, in order to understand better the effect of the interaction of superHubble modes with the DeSitter geometry and the (re-)appearance  of a quantum world during the reversal of the arrow of time near the turning point. We exhibited in \cite{richlaura1} that the phase space is not conserved due to these interactions. The equation that governs the evolution of $\rho$ is
\be
\label{eq:rhoevolve}
[\hat{H},\rho] = i\frac{d\rho}{dt}
\ee

After exit from inflation, during a matter dominated universe, the matter environment within our casual patch provides a strong supression in the off-diagonal elements of $\rho$ \cite{starobinski}, with width given by $\Omega_R^{-1}$,

\be
\rho \simeq e^{-\Omega_R a^4 (\phi - \phi')^2}
\ee
(Note that a large $\Omega_R$ means high squeezing in $\phi$  but a broad wavepacket in the conjugate momenta $p_{\phi}$ of $\phi$ ).
We can now obtain the evolution of $\rho$ by solving Eq.\ref{eq:rhoevolve} for late times, when the small vacuum energy $\Lambda$ starts dominating the total energy density in the Universe, $\hat{H}$,by including the effect of backreaction term. Solutions 
to Eq.\ref{eq:rhoevolve} give the following

\be
\label{eq:modifiedrho}
\rho \simeq e^{-[\frac{\Omega_R}{a^2} + \mu^2]a^6 (\phi-\phi')^2}
\ee

The second term is the correction coming from the backreaction correction, $S_n$, where $\mu^2 \simeq <\frac{n^2 a}{H_0}[ 1 - \frac{1}{3}(\frac{m^2 a}{n})^2 ]f_n^{2}>$. The first term is much studied in literature,(see e.g.\cite{starobinski}). For the sake of illustration, we can use as an approximate estimate for the first term $\Omega_R$ the results obtained in e.g \cite{decoherence}. Clearly, $\mu^2 \le 0$ for the backreaction of superhorizon modes $n \le \sqrt{2/3} aH_0$, on the DeSitter Universe. We can therefore see that the total Gaussian suppression in Eq.\ref{eq:modifiedrho}, $\xi = [\frac{\Omega_R}{a^2} + \mu^2]$ is decreasing and approaching zero fast as $\mu^2 \simeq \Sigma_n S_n f_n ^{2}$ becomes negative.

The reduced density matrix:
\bea
\rho &=& \int \Psi(a,\phi,f_n)\Psi(a',\phi',f'_n) \Pi_n df_n df'_n \nonumber \\
&\simeq& \rho_0 e^{-\frac{(a\pi)^6 H^2 \mu^2(\phi-\phi')^2}{2} }\nonumber \\
\rho_0 &\simeq& \langle\Psi_0 (a,\phi)\Psi_0(a'\phi')\rangle \nonumber \\
&\simeq& \rho(a,a') e^{-\xi a^6 (\phi-\phi')^2}.
\label{eq:density1}
\eea
Note that since $\mu^2 < 0$, we effectively have $\xi = \frac{\Omega_R}{a^2} - |\mu^2|$. As $\xi \rightarrow 0$ the classical trajectory in phase space $\phi$ is spreading thus losing its classicality and becoming highly quantum. The increasing width $\xi^{-1}$ of the Gaussian cross-term of the density matrix $\rho$ describes the braodening of the wavepacket during the transition to a quantum universe.  What effects would we observe as the universe is going toward the transition? This is a difficult question since near the turning point, the width of the Gaussian suppresion goes to infinity, thus the WKB approximation breaks down at the turning point. Hence, if we were to speculate as to what may we observe away from the turning point where the semiclassical treatment is still valid, as the broadening of the wavepacket is becoming comparable to the difference between neighboring energy levels $\phi, \phi'$, then we would probably expect to see an inhomogenious growth of anisotropies in the sky and probably different values of $\Lambda$ and quantum numbers in different parts of the sky, i.e. the opposite of the decoherence effect in the early universe \cite{zeh}. 

Varying the above action of matter+gravity, $S=S_g +
\Sigma_n S_{n,backreac}$ and using  the $\phi(a)$ trajectory equation of
motion, we can obtain the turning point for $\phi$ by the condition $\xi=0$ which gives
 $a_* \simeq \sqrt{\frac{2\Lambda}{3}}$ using the expressions above for $\mu$ and the time evolved solution $\rho$  .During a DeSitter phase in a time $a_*$, the
Gaussian width that suppresses the off-diagonal element of $\rho_r$ in
$\phi$ space is increasing due to the developing instability, $\rho_r
\simeq e^{-[\Omega_R - a^2\mu^2]a_{*}^4(\phi - \phi')^2}$.
The energy level in phase space is broadening in the landscape
coordinate $\phi$. By the uncertainty principle a broadening in $\phi$
means high squeezing in the space of its conjugate momenta $p_{\phi}$ since
$\Delta\phi \Delta p_{\phi} \simeq 1$. Thus the big crunch in real space
and the quantum recoherence in our universe as it goes through its turning
point in phase space. We can estimate the entropy of the DeSitter universe
as it is heading through its turning point by noting that the Euclidean
action is the entropy of the system. Roughly, $S\simeq
|r_{DeSitter} - r_{f_n}|^2 \simeq (\sqrt\frac{3}{\Lambda} -\frac{3a^2}{2 \Lambda^{3/2}})\rightarrow 0$. This means that the total
entropy for the DeSitter state including its horizon size
fluctuations goes to zero at $a_* \simeq O(\sqrt{\Lambda})$, \footnote{a result we could have expected physically,
for a quantum DeSitter state but that is confirmed here by a different calculation}, as the universe becomes quantum
through its turning point in phase space and big crunch in real space.
That is to say that the allowed volume in phase space for this classical
configuration shrinks to a point ie it is nearly zero. And so is the
dimension of the Hilbert space for it.This implies that there are no states available in the Hilbert space for a DeSitter universe.
 
\subsection{What have we learned}

The analysis carried in this work illustrates the fact that superhorizon nonlocal entanglement and the dynamics of gravity can not be ignored during a DeSitter era. When the DeSitter Universe loses causal contact with the internal environment,its only interaction left is with the higher multipoles. During this stage the DeSitter patch recoheres and ends up with its quantum entropy, which is zero, 
due to developing Jeans like instabilities. An immediate consequence of
this statement is that the arrow of time is reversed within a time $a_*$
during the DeSitter phase, due to the appearance of coherence and
broadening of levels, as the Universe heads towards its gravitationally induced turning point. Although the universe becomes quantum by being superimposed to its own reversed copy and other patches at its turning point, we as internal observers perceive this reversal as a catastrophic crunch in our patch.  The details at the turning point
depend on the boundary conditions imposed on $\dot\Psi(a_*)$. However at the turning point the quantum
universe is superimposed to its own time reversed copy and possibly other
patches such that its momenta in phase space squeezes to nearly zero. The
instability time when this phenomena occurs is, as we have shown here,$a_*
\simeq O(H_0)$. This timescale is exponentially shorter even than the tunneling time, if the metastable DeSitter space
could lower its energy by decaying. Tunneling to other vacua
may not help save the DeSitter Universe. Due to the quantum dynamics
of gravity this unstable state crunches catastrophically by becoming quantum during the time scales that it would have reached its final equlibrium and settle into a pure state. The mixing
does not allow DeSitter space to evolve and settle into a classical and pure equilibrium state with constant entropy at the end. Instead as the universe is trying to reach its DeSitter equilibrium, this entanglement induces a violent transition to a quantum universe for the final state of the DeSitter space. This recoherence occurs much before the universe can tunnel through to a lower energy configuration. As the DeSitter universe is becoming unstable, internal observers in this patch, perceive their thermodynamic arrow of time as being reversed during a timescale $a_*$ of order Hubble time.
Clearly we dont observe this in our universe. This fact in combination
with the result that the volume of phase space for DeSitter states is zero
leads us to conclude: Nature forbids the existence of a pure cosmological
constant. 
\section{Comments}
Then what are we to make of the observed acceleration of our universe?
Based on the above analysis, we would predict that this acceleration is a
temporary bleep due to some dynamic or transplanckian field and not our final destiny due to
a fundamental scale $\Lambda$ in nature. Observations will soon tell us
whether this is true or false. In our analysis we assume general relativity
as the theory of gravity and the validity of quantum mechanics. If it turns out that we
observe a pure $\Lambda$ then this must signal the breakdown of the general
relativity in the IR regime and the emergence of a new theory of gravity by
which our arguments above would not hold. It is though entirely possible
that, just like the second law of thermodynamics, we may have a new
empirical law in nature,let us call it the 'Entropy Exclusion Principle',
that is:{\it Nature Does Not Allow a Fundamental $\Lambda$}.

How can these predictions be tested? The combined observations from SN1a, large scale structure LSS and, CMB, from existing or upcoming experiments like Planck, SNAP and LISA, will soon be able to pin down whether the equation of state of dark energy is a constant or if it evolves with time\cite{melchiorri}. We concluded here that if it turns out we do observe a cosmological constant then gravity must be modified in the ultralarge scales, the IR regime. A modified theory of gravity would leave a strong signature on large scale structure of the universe\cite{marsn1a,hanestad}. Dark matter experiments e.g. \cite{darkmatter} and weak lensing tomography for LSS would reveal such deviations and test our predictions.

 It is interesting to note that the initial and boundary states are selected by the same superselection rule in phase space.
Without a precise knowledge of quantum gravity and the phase space for the wavefunctions, we can only speculate as to what happens next: if quantum mechanics remains valid deep into the quantum gravity realm then, after the turning point crunch, for example with a boundary condition $\dot\Psi=0$, the 'DeSitter' universe would travel its reversed path in the trajectory towards its initial point\footnote{For a different choice of boundary conditions for the wavepacket in phase space, the reversed path can be very different from the forward one depending on what WKB wavefunctions in the wavepacket dominate the action in the reversed direction, in which case there may not even be a cyclic behaviour at all. Subtleties related to the boundary conditions at the turning point were investigated by Hawking and Page \cite{noboundary} for the case of closed universes, pointed out to me by R.Holman. It would be interesting to investigate the role of the choice of boundary conditions for $\Psi(a_*)$ on the cosmological arrow of time,for DeSitter space but this issue is beyond the scope of the present paper.}, therefore completing one of its cycles between initial and final states. In this recycling the universe bounces between classical and quantum in each turning point, thus a classical world would be only an intermediate stage when the universe is away from the turning point in its trajectory. Near the turning point the wavepacket would spread out by disintegrating into its many components and thus the Universe would behave as a quantum system. However decoherence in each cycle takes its toll: each cycle gets shorter and shorter due to the shifting of the energies by the interaction with states. (The shifting of the energy is a well known result in particle physics when a particle interacts with a field than its trajectory is modified due to its energy shift by interaction\cite{decoherence}.). It is possible in this picture that after many such cycles, with shorter and shorter trajectories, the universe will spiral in to one point and disappear. Could this speculation provide us with a mechanism for DeSitter space evaporating away its $\Lambda$ during its recycling in phase space?

\begin{acknowledgments} I would like to thank R. Holman for valuable discussions and comments. LMH was supported in part by DOE grant DE-FG02-06ER41418 and NSF grant PHY-0553312.
\end{acknowledgments}

\section{Appendix}
Here we summarize the calculation for the backreaction potential term $U_{-}(\alpha,\phi)$, of Eq.\ref{eq:master}. Based on Ref.\cite{kiefer2,hh,richlaura1}, a time parameter $t$ is defined for WKB wavefunctions of the universe such that the equation for the perturbation modes $\psi_n$ in the wavefunction can be written as a Schr$\ddot{\rm o}$dinger equation and be consistent with Einstein equations. 
If $S$ is the action for the mean values $\alpha, \phi$ for a nearly DeSitter state with scale factor $a=Log[[\alpha]$ and vacuum energy given roughly by $\Lambda \simeq 1/2 (m^2 \phi^2)$, then by defining the parameter $y \equiv \left(\partial S\slash \partial \alpha\right)\slash \left(\partial S\slash \partial \phi\right)\sim \dot{\alpha}\slash \dot{\phi}$, one can write in the second order WKB approximation, \cite{hh,kiefer2}:
\bea
\label{eq:pertschr}
\psi_n &=& e^{\frac{\alpha}{2}} \exp\left(i \frac{3}{2 y} \frac{\partial S}{\partial \phi} f_n^2\right)\psi_n^{(0)}\nonumber \\
i\frac{\partial \psi_n^{(0)}}{\partial t}  &=& e^{-3 \alpha} \left\{-\frac{1}{2} \frac{\partial^2}{\partial f_n^2} + U(\alpha,\phi) f_n^2\right\} \psi_n^{(0)}\nonumber \\
U(\alpha,\phi) &=&e^{6\alpha} \left\{(\frac{n^2-1}{2})e^{-2\alpha} +\frac{m^2}{2} +\right .\nonumber \\
&+& \left . 9m^2 y^{-2}\phi^2
-6m^2 y^{-1}\phi\right\}.
\eea

During the nearly DeSitter stage, $S\approx-1\slash 3\ m e^{3\alpha} \phi_{\rm inf} \approx -1\slash 3 e^{3\alpha} \sqrt{\Lambda}$, where $\phi_{\rm inf}$ is the value of the field during inflation, so that $y=3\phi_{\rm inf}$. Thus long wavelength matter fluctuations are amplified during a vacuum energy dominated universe, and driven away from their ground state \cite{hh}.This is not the case for a matter or radiation dominated universe.When the universe exits the DeSitter stage, the wavepacket for the universe is in an oscillatory regime, $y$ is large so that the potential $U(\alpha, \phi)$ changes from $U_{-}(\alpha, \phi)$ to $U_{+}(\alpha, \phi)$, where 
\be
\label{eq:posnegU}
U_{\pm} (\alpha, \phi) \sim e^{6\alpha} \left[\frac{n^2-1}{2}e^{-2\alpha}\pm \frac{m^2}{2}\right].
\ee
as can be seen from Eq.(\ref{eq:pertschr}). It is straightforward to check from the equation of motion for the superHubble perturbation modes that in this case, $U=U_{+}$, as it was shown in \cite{hh}, these modes are frozen in as in the conventional theory of perturbations. We can follow the evolution of these superHubble matter perturbation during the second stage of a vacuum energy domination phase, by using the Schrodinger equation above. During the nearly DeSitter state in the future, these modes develop gravitational instabilities because for reasons shown above, again $U(\alpha \phi)<0$.The matter perturbation modes $f_n$ with $n \le aH_0$ are coupled to the background gravitational potential, therefore the main difference with the mechanism described in \cite{richlaura1} for the instabilities during an early DeSitter stage, is that the scale of the curvature at late times, is of the order the low energy DeSitter Hubble horizon, $m^2 \simeq \sqrt{\Lambda}$. From Eq.(\ref{eq:pertschr}) we see that during a (nearly) DeSitter stage, the patches that have $U(\alpha, \phi)<0$, which can happen for small enough physical wave vector $k_n = n e^{-\alpha}$, develop tachyonic instabilities due to the growth of perturbations: $\psi_n \simeq e^{-\mu_n \alpha} e^{i \mu_n \phi}$, where $-\mu_n^2 = U(\alpha,\phi)f_n^2$. Thus the solution for the wavefunction of the universe, in phase space given by $\Psi(a,\phi,f_n)=\Psi_0(a,\phi)\Pi_n \psi_n(a,\phi,f_n)$ is damped in the intrinsic time $\alpha$ rather than oscillatory. The damping of these wavefunctions is correlated with the tachyonic, Jeans-like instabilities that develop in real space-time for the corresponding matter perturbation modes, consistent with Einstein equations for matter perturbation $f_n$; when $U(\alpha, \phi)<0$, $f_n\sim e^{\pm \mu_n t}$, while for $U(\alpha, \phi)>0$, the long wavelength matter perturbations $f_n$ are frozen in. 
The gravitational instabilities of matter fluctuation in real spacetime, can be seen from the equation of motion for $\phi,f_n$ obtained by varying the action with respect to these variables. For the tachyonic case $U<0$ universes, we have

\begin{equation}
\ddot{f}_n + 3H\dot{f}_n + \frac{U_{-}}{a^3} f_n =0,
\label{eq:modegrowth}
\end{equation}

From Eqn.\ref{eq:posnegU} we can see that only during a DeSitter stage, when $U<0$ one obtains growing and decaying solution in spacetime roughly  for $f_n \simeq e^{\pm \mu t}$ and the growth of instabilites in the infinite number of superHubble matter modes $f_n$ with $0<n<aH_0$ which can eventually overcome the pressure of the vacuum energy on spacetime and drive the wavefunction of the universe away from its classical trajectory.

\end{document}